\documentclass[10pt,conference]{IEEEtran}
\IEEEoverridecommandlockouts

\ifCLASSOPTIONcompsoc
\usepackage[caption=false,font=normalsize,labelfon
t=sf,textfont=sf]{subfig}
\else
\usepackage[caption=false,font=footnotesize]{subfi
g}
\fi

\usepackage{amsmath,amssymb}
\usepackage{multicol}
\usepackage{graphicx,graphics,color,psfrag}
\usepackage{cite,balance}
\usepackage{algorithm}
\usepackage{accents}

\usepackage{bm}
\usepackage{url}
\usepackage{algorithmic}
\usepackage[english]{babel}
\usepackage{multirow}
\usepackage{enumerate}
\usepackage{cases}
\usepackage{stfloats}
\usepackage{dsfont}
\usepackage{color,soul}
\usepackage{amsfonts}
\usepackage{tcolorbox}
\usepackage{amsmath}
\usepackage{float}

\usepackage{cite,graphicx,amsmath,amssymb}
\usepackage{fancyhdr}
\usepackage{hhline}
\usepackage{graphicx,graphics}
\usepackage{array,color}
\usepackage{amsmath}
\usepackage{stfloats}
\usepackage[flushleft]{threeparttable}
\usepackage{booktabs}

\newtheorem{proposition}{Proposition}
\newtheorem{remark}{Remark}

\ifCLASSINFOpdf

\else

\fi

\begin{document}
\title{Fully-Passive versus Semi-Passive IRS-Enabled Sensing: SNR Analysis}

\author{
\IEEEauthorblockN{Xianxin~Song\IEEEauthorrefmark{1},  Xinmin~Li\IEEEauthorrefmark{1}\IEEEauthorrefmark{2}, Xiaoqi~Qin\IEEEauthorrefmark{3}, and~Jie~Xu\IEEEauthorrefmark{1}
\thanks{Xinmin~Li and Jie Xu are the corresponding authors.}
}

\IEEEauthorblockA{\IEEEauthorrefmark{1}School of Science and Engineering (SSE) and Future Network of Intelligence Institute (FNii),\\ The Chinese University of Hong Kong (Shenzhen), Shenzhen,  China}
\IEEEauthorblockA{\IEEEauthorrefmark{2}The School of Information Engineering, Southwest University of Science and Technology, Mianyang, China}
\IEEEauthorblockA{\IEEEauthorrefmark{3}The State Key Laboratory of Networking and Switching Technology, \\ Beijing University of Posts and Telecommunications, Beijing, China}
Email: xianxinsong@link.cuhk.edu.cn, lixm@swust.edu.cn, xiaoqiqin@bupt.edu.cn, xujie@cuhk.edu.cn
}

\maketitle
\begin{abstract}
This paper compares the signal-to-noise ratio (SNR) performance between the fully-passive intelligent reflecting surface (IRS)-enabled non-line-of-sight (NLoS) sensing versus its semi-passive counterpart. In particular, we consider a basic setup with one base station (BS), one uniform linear array (ULA) IRS, and one point target at the BS's NLoS region, in which the BS and the IRS jointly design the transmit and reflective beamforming for performance optimization. By considering two special cases with the BS-IRS channels being line-of-sight (LoS) and Rayleigh fading, respectively, we derive the corresponding asymptotic sensing SNR when the number of reflecting elements $N$ at the IRS becomes sufficiently large. It is revealed that in the two special cases, the sensing SNR increases proportional to $N^2$ for the semi-passive IRS sensing system, but proportional to $N^4$ for the fully-passive IRS sensing system. As such, the fully-passive IRS sensing system is shown to outperform the semi-passive counterpart when $N$ becomes large, which is due to the fact that the fully-passive IRS sensing enjoys  additional reflective beamforming gain from the IRS to the BS that outweighs the resultant path loss in this case. Finally, numerical results are presented to validate our analysis under different transmit and reflective beamforming design schemes.
\end{abstract}


\IEEEpeerreviewmaketitle

\section{Introduce}
Integrated sensing and communication (ISAC) has been recognized as one of the key usage scenarios for future sixth-generation (6G) networks\cite{IMT2030}, in which cellular base stations (BSs) are enabled to utilize radio signals for wireless (radar) sensing\cite{liu2021integrated,9540344}. In general, wireless sensing relies on the line-of-sight (LoS) links between the BS transceiver and sensing targets\cite{richards2014fundamentals}, thus making the non-LoS (NLoS) target sensing at the LoS-blocked areas of BSs a challenging task. 

Intelligent reflecting surface (IRS)\cite{8811733,9326394} or reconfigurable intelligent surface (RIS)\cite{9424177} has become a promising solution to enable NLoS wireless sensing\cite{Stefano,xianxin,sankar2022beamforming,song2021joint,xianxin2023ICC,9724202,wang2022stars,fangyuan}. In particular, by reconfiguring wireless propagation environments, IRS can provide reflected virtual LoS links to bypass environment obstructions for extending the sensing coverage. Furthermore, by properly adjusting the phase shifts of reflecting elements, IRS can form reflective beamforming to enhance the reflected signal strength for facilitating sensing and communications.


In general, the IRS-enabled sensing can be implemented based on two different structures with fully-passive \cite{Stefano,xianxin,sankar2022beamforming,song2021joint,xianxin2023ICC} and semi-passive IRSs\cite{9724202,wang2022stars,fangyuan}, respectively, depending on whether the IRS is deployed with dedicated sensors for receiving and processing target echo signals. For the fully-passive IRS without dedicated sensors, the target sensing is performed at the BS based on the echo signals through the BS-IRS-target-IRS-BS link. By contrast, for the semi-passive IRS with dedicated sensors, the target sensing is directly performed at the IRS based on the echo signals through the BS-IRS-target-IRS link. In the literature, the authors in \cite{Stefano} and  \cite{xianxin}  investigated the joint transmit and reflective beamforming design for the fully-passive IRS-enabled target detection and estimation, respectively, while \cite{sankar2022beamforming,song2021joint,xianxin2023ICC} studied the joint beamforming design for the fully-passive IRS-enabled ISAC. On the other hand, prior work \cite{9724202} studied the reflective beamforming optimization for the semi-passive IRS-enabled sensing, and \cite{wang2022stars} \cite{fangyuan} investigated the joint beamforming design for the semi-passive IRS-enabled ISAC.

The fully-passive and semi-passive IRS sensing systems have their pros and cons. As compared to the semi-passive counterpart, the received echo signals in the fully-passive IRS sensing are subject to one more signal reflection from the IRS to the BS. This not only results in additional path loss that is harmful for sensing, but also leads to additional reflective beamforming gain that is beneficial for sensing. By combining the above drawback and benefit, an interesting questions arises: Under what conditions does the fully-passive IRS sensing outperform the semi-passive one? This thus motivates us to analyze and compare the sensing performance of the fully-passive and semi-passive IRS sensing systems in this work.

In particular, this paper considers a basic IRS-enabled sensing setup with one BS, one uniform linear array (ULA) IRS, and one point target at the NLoS region of the BS, in which the BS and the IRS can jointly optimize the transmit and reflective beamforming designs for performance optimization. Under this setup, we compare the sensing signal-to-noise ratio (SNR) performance with fully-passive IRS versus that with semi-passive IRS. By considering two special channel models for the BS-IRS links, namely the LoS and Rayleigh fading channels, we derive the asymptotic sensing SNRs with the fully-passive and semi-passive IRSs when the number of reflection elements $N$ at the IRS becoming sufficiently large. It is shown that for the semi-passive IRS sensing, the asymptotic sensing SNR increases proportional to $N^2$ due to the reflecting beamforming over the forward link from the BS to the IRS, while for the fully-passive IRS sensing, the asymptotic SNR is proportional to $N^4$ thanks to the additional reflecting beamforming over the backward link from the IRS to the BS. As a result, the sensing SNR with the fully-passive IRS sensing outperforms the semi-passive counterpart when $N$ is greater than a certain threshold, as the additional beamforming gain over the backward link is larger then the corresponding path loss in this case. Finally, numerical results are presented to validate our analysis under different transmit and reflective beamforming design schemes.

\textit{Notations:} 
Boldface letters refer to vectors (lower case) or matrices (upper case). For a square matrix $\mathbf S$, $\mathrm {tr}(\mathbf S)$ denotes its trace, and $\mathbf S \succeq \mathbf{0}$ means that $\mathbf S$ is positive semi-definite. For an arbitrary-size matrix $\mathbf M$, $\mathbf M^{T}$ and $\mathbf M^{H}$ denote its transpose and conjugate transpose, respectively. We use $\mathcal{C N}(\mathbf{0}, \mathbf{\Sigma})$ to denote the distribution of a circularly symmetric complex Gaussian (CSCG) random vector with zero mean $\mathbf 0$ and covariance matrix $\mathbf \Sigma$, and $\sim$ to denote “distributed as”. The spaces of $x \times y$ complex matrix is denoted by $\mathbb{C}^{x \times y}$. 
The symbol $\|\cdot\|$ denotes the Euclidean norm, $|\cdot|$ denotes the magnitude of a complex number, and $\mathrm {diag}(a_1,\cdots,a_N)$ denotes a diagonal matrix with diagonal elements $a_1,\cdots,a_N$.

\section{System Model}
\begin{figure}[t]
		\centering
        \includegraphics[width=0.45\textwidth]{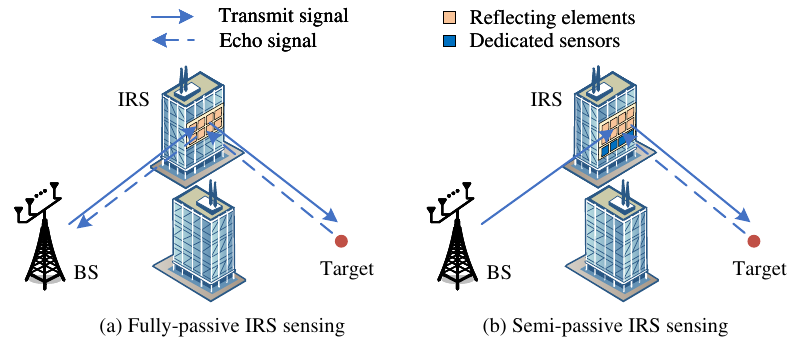}
    	\caption{System model of IRS-enabled sensing.}
    	\label{system_model}
    	\vspace{-10pt}
\end{figure}

We consider an IRS-enabled NLoS target sensing system as shown in Fig.~\ref{system_model}, which consists of a BS with $M_t$ transmit and $M_r$ receive antennas, a ULA IRS with $N$ reflecting elements, and a point target at the NLoS region of the BS. In particular, we consider two IRS architectures, namely fully-passive and semi-passive IRSs, which are deployed without and with dedicated sensors as shown in Fig. 1(a) and Fig. 1(b), respectively. For fair comparison, we assume that there are $M_r$ dedicated sensors or receive antennas at the semi-passive IRS for sensing.

We consider one particular sensing block with $T$ symbols. Let $\mathbf x(t)$ denote the transmitted sensing signal by the BS at symbol $t \in \mathcal T = \{1,\cdots, T\}$. The sample covariance matrix of the transmitted signal is denoted by
$\mathbf R = \frac{1}{T}\sum_{t\in \mathcal{T}} \mathbf x(t) \mathbf x^{H}(t) \succeq \mathbf 0$.
Let $P_0$ denote the maximum transmit power at the BS. Then we have $\frac{1}{T}\sum_{t\in \mathcal T}\| \mathbf x(t) \|^2 =\mathrm {tr}(\mathbf R) \le P_0$.
Let $\mathbf{G}_t \in \mathbb{C}^{N \times M_t}$ and $\mathbf{G}_r \in \mathbb{C}^{M_r \times N}$ denote the channel matrices of the BS-IRS and IRS-BS links, respectively. 
Let $\mathbf{\Phi} = \mathrm {diag}(e^{j\phi_1},\cdots,e^{j\phi_{N}})$ denote the reflection matrix of the IRS, with $\phi_n \in (0, 2\pi]$ denoting the phase shift of reflecting element $n \in \mathcal N  = \{1,\cdots,N\}$. 

First, we consider the fully-passive IRS sensing in Fig. 1(a). Without dedicated sensors at the IRS, the NLoS target sensing is performed at the BS through the BS-IRS-target-IRS-BS link. Let $\theta$ denote the direction-of-arrival (DoA) of the target with respect to (w.r.t.) the IRS. Let $\mathbf a (\theta)$ denote the steering vector at the IRS with angle $\theta$. By choosing the center of the ULA antennas as the reference point\cite{bekkerman2006target}, the steering vector $\mathbf a(\theta)$ is expressed as
\begin{equation}
\mathbf a (\theta) = \left[e^{-\frac{ j\pi(N-1)\hat{d}\sin \theta}{\lambda}},e^{\frac{-j\pi(N-3)\hat{d}\sin \theta}{\lambda}},\cdots,e^{\frac{ j\pi(N-1)\hat{d}\sin \theta}{\lambda}}\right]^{T},
\end{equation}
where $\hat{d}$ denotes the spacing between consecutive reflecting elements at the IRS, and $\lambda$ denotes the carrier wavelength.  The received echo signal by the BS through the BS-IRS-target-IRS-BS link at symbol $t \in \mathcal{T}$ is
\begin{equation}\label{eq:echo_signal_I}
\mathbf y_1(t)= \alpha\mathbf{G}_r\mathbf{\Phi}^T\mathbf a (\theta)\mathbf a^{T}(\theta) \mathbf{\Phi}\mathbf{G}_t \mathbf x(t)+ \mathbf n_1(t),
\end{equation}
where $\mathbf n_1(t)\sim \mathcal{C N}(\mathbf{0}, \sigma^2\mathbf I_{M_r})$ denotes the additive white Gaussian noise (AWGN) at the BS receiver, and $\alpha \in \mathbb{C}$ denotes the channel coefficient of the IRS-target-IRS link that depends on both the target radar cross section (RCS) and the round-trip path loss of the IRS-target-IRS link. Accordingly, the BS performs the target sensing based on the received signal $\mathbf y_1(t)$ in \eqref{eq:echo_signal_I}. 

Next, we consider the semi-passive IRS sensing in Fig. 1(b). With dedicated sensors at the IRS, the NLoS target sensing is directly performed at the IRS based on the received signal through the BS-IRS-target-IRS link. For fair comparison, we assume that the the spacing between consecutive sensors at the IRS is also $\hat{d}$. With target angle $\theta$, the steering vector at the sensors of IRS is denoted as
\begin{equation}
\mathbf b (\theta)\! =\! \left[e^{-\frac{ j\pi(M_r-1)\hat{d}\sin \theta}{\lambda}}\!,e^{\frac{-j\pi(M_r-3)\hat{d}\sin \theta}{\lambda}}\!,\cdots,e^{\frac{ j\pi(M_r-1)\hat{d}\sin \theta}{\lambda}}\right]^{T}.
\end{equation}
In this case, the received echo signal by the IRS through the BS-IRS-target-IRS link at symbol $t \in \mathcal{T}$ is
\begin{equation}\label{eq:echo_signal_II}
\mathbf y_2(t)=\alpha\mathbf b(\theta)\mathbf a^{T}(\theta) \mathbf{\Phi}\mathbf{G}_t\mathbf x(t)+ \mathbf n_2(t),
\end{equation}
where $\mathbf n_2(t) \sim \mathcal{C N}(\mathbf{0}, \sigma^2\mathbf I_{M_r})$ denotes the AWGN at the IRS receivers. Accordingly, the semi-passive IRS performs the target sensing based on the received signal $\mathbf y_2(t)$ in \eqref{eq:echo_signal_II}. For notational convenience, in the sequel we drop $\theta$ in $\mathbf a(\theta)$ and $\mathbf b(\theta)$ and accordingly denote them as $\mathbf a $ and $\mathbf b$, respectively.



Then, we consider the sensing SNR as the performance metric for the above two IRS sensing systems, similarly as in prior work \cite{richards2014fundamentals,Stefano}, which is valid for target detection tasks.\footnote{The target detection probability is generally a monotonically increasing function w.r.t. the sensing SNR when the target is present \cite{richards2014fundamentals,Stefano}.} Based on the received echo signals $\mathbf y_1(t)$ in \eqref{eq:echo_signal_I} and  $\mathbf y_2(t)$ in \eqref{eq:echo_signal_II}, the sensing SNRs of the fully-passive and semi-passive IRS sensing systems are respectively given by
\begin{equation}\label{eq:SNR_fully-passive}
\begin{split}
\text{SNR}_1 (\mathbf R, \mathbf \Phi) =& \frac{\mathbb{E}(\| \alpha\mathbf{G}_r\mathbf{\Phi}^T\mathbf a\mathbf a^{T} \mathbf{\Phi}\mathbf{G}_t\mathbf x(t)\|^2)}{\sigma^2}\\
 =&\frac{|\alpha|^2\|\mathbf{G}_r\mathbf{\Phi}^{T}\mathbf a\|^2\mathbf a^{T} \mathbf{\Phi}\mathbf{G}_t\mathbf R \mathbf{G}_t^{H}\mathbf{\Phi}^{H}\mathbf a^*}{\sigma^2},
 \end{split}
\end{equation}
\begin{equation}\label{eq:SNR_semi-passive}
\begin{split}
\text{SNR}_2 (\mathbf R, \mathbf \Phi) =& \frac{\mathbb{E}(\| \alpha\mathbf b\mathbf a^{T} \mathbf{\Phi}\mathbf{G}\mathbf x(t)\|^2)}{\sigma^2}\\
 =&\frac{|\alpha|^2\|\mathbf b\|^2\mathbf a^{T} \mathbf{\Phi}\mathbf{G}_t\mathbf R \mathbf{G}_t^{H}\mathbf{\Phi}^{H}\mathbf a^*}{\sigma^2}.
 \end{split}
\end{equation}
By comparing the numerators of \eqref{eq:SNR_fully-passive} and \eqref{eq:SNR_semi-passive}, the fully-passive and semi-passive IRS sensing systems have the same transmit beampatterns of $\mathbf a^{T} \mathbf{\Phi}\mathbf{G}_t\mathbf R \mathbf{G}_t^{H}\mathbf{\Phi}^{H}\mathbf a^*$, but different receive beampatterns, i.e., $\|\mathbf{G}_r\mathbf{\Phi}^{T}\mathbf a\|^2$ versus $\|\mathbf b\|^2$, respectively.
By comparing the receive beampatterns, it is clear that the fully-passive IRS sensing system experiences additional path loss due to the  additionally multiplied channel matrix $\mathbf G_r$ in $\|\mathbf G_r \mathbf \Phi^T \mathbf a\|^2$, but enjoys stronger reflective beamforming gain thanks to the additional reflective beamformer $\mathbf \Phi$ therein. This thus introduces an interesting tradeoff in balancing between the path loss versus the reflective beamforming gain. In the following, we compare the SNR performance by considering that the BS and the IRS can jointly optimize the transmit beamforming (or the sample covariance matrix $\mathbf R$) and the reflective beamforming (or the reflection matrix $\mathbf \Phi$).



\section{Sensing SNR Analysis}
This section analyzes the sensing SNR performance of the fully-passive and semi-passive IRS sensing systems, in which the transmit and reflective beamformers are jointly optimized. In the following, Section III-A first focuses on the SNR with transmit beamforming optimization only, and Sections III-B and III-C consider the SNRs with joint beamforming optimization under two special cases with LoS and Rayleigh fading channels for the BS-IRS links, respectively. 

\subsection{Sensing SNR Comparison with Transmit Beamforming Optimization Only}
First, we consider the sensing SNR with only transmit beamforming optimization. It is well established that the maximum ratio transmission (MRT) is optimal to maximize $\mathbf a^{T} \mathbf{\Phi}\mathbf{G}_t\mathbf R \mathbf{G}_t^{H}\mathbf{\Phi}^{H}\mathbf a^*$ or equivalently maximize the sensing SNRs in \eqref{eq:SNR_fully-passive} and \eqref{eq:SNR_semi-passive} for both fully-passive and semi-passive IRS sensing, i.e.,
\begin{equation}\label{eq:MRT}
\mathbf R_x^\text{MRT} =\frac{P_0\mathbf G_t^{H} \mathbf \Phi^H  \mathbf a^* \mathbf a^{T} \mathbf \Phi \mathbf G_t}{\|\mathbf G_t^{T} \mathbf \Phi^T \mathbf a\|^2}.
\end{equation}

By substituting $\mathbf R_x^\text{MRT}$ into \eqref{eq:SNR_fully-passive} and \eqref{eq:SNR_semi-passive}, the resultant sensing SNRs with transmit beamforming optimization for  fully-passive and semi-passive IRS sensing are respectively given by
\begin{equation}\label{eq:SNR_I_opt}
\widehat{\text{SNR}}_1(\mathbf{\Phi}) =\frac{P_0|\alpha|^2\|\mathbf{G}_r\mathbf{\Phi}^T\mathbf a\|^2\|\mathbf{G}_t^T\mathbf{\Phi}^T\mathbf a\|^2}{\sigma^2},
\end{equation}
\begin{equation}\label{eq:SNR_II_opt}
\widehat{\text{SNR}}_2(\mathbf{\Phi}) =\frac{P_0|\alpha|^2\|\mathbf b\|^2\|\mathbf{G}_t^T\mathbf{\Phi}^T\mathbf a\|^2}{\sigma^2}.
\end{equation}
Based on the SNRs in \eqref{eq:SNR_fully-passive}, \eqref{eq:SNR_semi-passive}, \eqref{eq:SNR_I_opt}, and \eqref{eq:SNR_II_opt}, we directly have the following proposition, for which the proof is omitted.
\begin{proposition} \label{lemma: SNR}
Under any given or optimized transmit beamforming, the sensing SNR of the fully-passive IRS sensing is greater than that of the semi-passive IRS sensing (i.e., $\text{SNR}_1 (\mathbf R, \mathbf \Phi) > \text{SNR}_2 (\mathbf R, \mathbf \Phi)$ and $\widehat{\text{SNR}}_1(\mathbf{\Phi}) > \widehat{\text{SNR}}_2(\mathbf{\Phi})$) when
\begin{equation}
  \|\mathbf{G}_r\mathbf{\Phi}^{T}\mathbf a\|^2> \|\mathbf b\|^2=M_r.
\end{equation}
\end{proposition}

Proposition~\ref{lemma: SNR} shows that reflective beamforming $\mathbf \Phi$ is a critical factor that affects the sensing SNR performance for IRS sensing. It is expected that when the number of reflecting elements $N$ at IRS or the dimension of $\mathbf \Phi$ becomes large and with properly optimized $\mathbf \Phi$, the fully-passive IRS sensing may outperform the semi-passive counterpart thanks to the additional reflective beamforming gain. 

\subsection{Asymptotic Sensing SNR Performance with Joint Beamforming Optimization under  LoS Channel} 
This subsection considers the special case when the BS-IRS channels are LoS. Accordingly, we analyze the asymptotic sensing SNR performance with joint beamforming optimization (i.e., the reflective beamforming optimization is further employed in addition to the MRT in Section III-A). In this case, the BS-IRS and IRS-BS channel matrices are given by
\begin{equation}
\mathbf G_t = \sqrt{L(d)} \mathbf h \mathbf g_t^T,
\mathbf G_r = \sqrt{L(d)} \mathbf g_r \mathbf h^T,
\end{equation} 
where $L(d)$ denotes the distance-dependent path loss, and $\mathbf h\in \mathbb{C}^{N \times 1}$, $\mathbf g_t\in \mathbb{C}^{M_t \times 1}$, and $\mathbf g_r\in \mathbb{C}^{M_r \times 1}$ correspond to the steering vectors at IRS reflecting elements, the transmitters of the BS, and the receivers of the BS, respectively. We thus have 
\begin{equation}\label{eq:LoS_G_t}
\|\mathbf{G}_t^T\!\mathbf{\Phi}^T\!\mathbf a\|^2\!=\!L(d)\|\mathbf g_t\|^2 |\mathbf h^T\!\mathbf{\Phi}^T\!\mathbf a|^2\!=\!L(d)M_t |\mathbf h^T\!\mathbf{\Phi}^T\!\mathbf a|^2,
\end{equation} 
\begin{equation}\label{eq:LoS_G_r}
\|\mathbf{G}_r\!\mathbf{\Phi}^T\!\mathbf a\|^2\!=\!L(d)\|\mathbf g_r\|^2 |\mathbf h^T\!\mathbf{\Phi}^T\!\mathbf a|^2\!=\!L(d)M_r |\mathbf h^T\!\mathbf{\Phi}^T\!\mathbf a|^2.
\end{equation}
By substituting \eqref{eq:LoS_G_t} and \eqref{eq:LoS_G_r} into the SNRs with transmit beamforming optimization in \eqref{eq:SNR_I_opt} and \eqref{eq:SNR_II_opt}, we have 
\begin{equation}\label{eq:SNR_I_opt_LoS}
\widetilde{\text{SNR}}_1(\mathbf{\Phi}) =\frac{P_0|\alpha|^2L^2(d)M_tM_r\|\mathbf h^T\mathbf{\Phi}^T\mathbf a\|^4}{\sigma^2},
\end{equation}
\begin{equation}\label{eq:SNR_II_opt_LoS}
\widetilde{\text{SNR}}_2(\mathbf{\Phi}) =\frac{P_0|\alpha|^2L(d)M_tM_r\|\mathbf{h}^T\mathbf{\Phi}^T\mathbf a\|^2}{\sigma^2}.
\end{equation}
Based on \eqref{eq:SNR_I_opt_LoS} and \eqref{eq:SNR_II_opt_LoS}, maximizing the sensing SNR is equivalent to maximizing $\|\mathbf h^T\mathbf{\Phi}^T\mathbf a\|^2$ for both fully-passive and semi-passive IRS sensing, i.e.,
\begin{subequations}
  \begin{align}\notag
    \text{(P1)}:\max_{\mathbf \Phi}&\quad  \|\mathbf h^T\mathbf{\Phi}^T\mathbf a\|^2\\ 
    \text { s.t. }& \quad  |\mathbf \Phi_{n,n}|=1, \forall n\in \mathcal{N}.
  \end{align}
\end{subequations}
According to the Cauchy-Schwarz inequality, the optimal solution of (P1) is  \
\begin{equation}\label{eq:RF_LoS}
\mathbf{\Phi}^\star = \mathrm {diag}(e^{j\phi_1^\star},\ldots,e^{j\phi_{N}^\star}),
\end{equation}
 where $\phi_n^\star = -\mathrm{arg}(h_n) -\mathrm{arg}(a_n), \forall n \in \mathcal{N}$, with $h_n$ and $a_n$ being the $n$-th element of vectors $\mathbf h$ and $\mathbf a$, respectively. 
With the optimal reflective beamformer $\mathbf \Phi^\star$ in \eqref{eq:RF_LoS}, we have
\begin{equation}\label{eq:h_phi_a}
\|\mathbf h^T\mathbf (\mathbf \Phi^\star)^T\mathbf a\|^2 = N^2. 
\end{equation}

By substituting \eqref{eq:h_phi_a} into the SNRs in \eqref{eq:SNR_I_opt_LoS} and \eqref{eq:SNR_II_opt_LoS}, we have the following proposition.
\begin{proposition}\label{prop:LoS}
For the case with LoS channel and with optimal joint beamforming, the resultant sesning SNRs of the fully-passive and semi-passive IRS sensing systems become 
\begin{equation}
\text{SNR}_1^\star=\frac{P_0|\alpha|^2L(d)^2M_tM_r N^4}{\sigma^2},
\end{equation}
\begin{equation}
\text{SNR}_2^\star =\frac{P_0|\alpha|^2L(d)M_tM_r N^2}{\sigma^2},
\end{equation}
which increase proportional to $N^4$ and $N^2$, respectively.
\end{proposition}

Based on Propositions~\ref{prop:LoS}, we compare the sensing SNR with joint beamforming optimization for the fully-passive and semi-passive IRS sensing systems in the following proposition.
\begin{proposition} \label{lemma_LoS}
For the case with LoS channel and with optimal joint beamforming, the sensing SNR of the fully-passive IRS sensing is greater than that of the semi-passive IRS sensing (i.e., $\widetilde{\text{SNR}}_1(\mathbf{\Phi}^\star)>\widetilde{\text{SNR}}_2(\mathbf{\Phi}^\star)$) when 
\begin{equation}
N>\frac{1}{\sqrt{L(d)}}.
\end{equation}
\end{proposition}
\subsection{Asymptotic Sensing SNR Performance with Joint Beamforming Optimization under  Rayleigh Fading Channel} 
This subsection considers another special case when the BS-IRS channels follow Rayleigh fading. Accordingly, we analyze the asymptotic sensing SNR performance with optimal joint beamforming. In this case, the BS-IRS and IRS-BS channel matrices are respectively given by 
\begin{equation}\label{eq:Rayleigh_G_t}
\mathbf G_t = \sqrt{L(d)} \hat{\mathbf G}_t,
\mathbf G_r = \sqrt{L(d)} \hat{\mathbf G}_r,
\end{equation} 
where $\hat{\mathbf G}_t \in \mathbb{C}^{N\times M_t}$ and $\hat{\mathbf G}_r \in \mathbb{C}^{M_r\times N}$ are CSCG random matrices with zero mean and unit variance for each element.
By substituting \eqref{eq:Rayleigh_G_t} into the SNRs with transmit beamforming optimization in \eqref{eq:SNR_I_opt} and \eqref{eq:SNR_II_opt}, we have 
\begin{equation}\label{eq:SNR_I_opt_Rayleigh}
\overline{\text{SNR}}_1(\mathbf{\Phi}) =\frac{P_0|\alpha|^2L^2(d)\|\hat{\mathbf G}_t^T\mathbf{\Phi}^T\mathbf a\|^2\|\hat{\mathbf G}_r\mathbf{\Phi}^T\mathbf a\|^2}{\sigma^2},
\end{equation}
\begin{equation}\label{eq:SNR_II_opt_Rayleigh}
\overline{\text{SNR}}_2(\mathbf{\Phi}) =\frac{P_0|\alpha|^2L(d)\|\hat{\mathbf G}_r\mathbf{\Phi}^T\mathbf a\|^2}{\sigma^2}.
\end{equation}

First, we consider the reflective beamforming with fully-passive IRS. In this case, maximizing $\overline{\text{SNR}}_1(\mathbf{\Phi})$ is equivalent to maximizing $\|\hat{\mathbf G}_t^T\mathbf{\Phi}^T\mathbf a\|^2\|\hat{\mathbf G}_r\mathbf{\Phi}^T\mathbf a\|^2$, which is formulated as
\begin{subequations}
  \begin{align}\notag
    \text{(P2)}:\max_{\mathbf \Phi}&\quad  \|\hat{\mathbf G}_t^T\mathbf{\Phi}^T\mathbf a\|^2\|\hat{\mathbf G}_r\mathbf{\Phi}^T\mathbf a\|^2\\ \label{eq:unit-modulus} 
    \text { s.t. }& \quad  |\mathbf \Phi_{n,n}|=1, \forall n\in \mathcal{N}.
  \end{align}
\end{subequations}
Problem (P2) is non-convex due to the non-convexity of the objective function and  the unit-modulus constraint in \eqref{eq:unit-modulus}. Though it is generally difficult to find its optimal solution, we can find a high-quality suboptimal solution by using the techniques of semi-definite relaxation (SDR) and successive convex approximation (SCA). The detailed algorithm is given in Appendix~\ref{sub:algorithm_for_p2}.

Based on the optimization in (P2), we analyze the resultant sensing SNR performance with fully-passive IRS. Let $\gamma^\star_1$ denote the optimal value of problem (P2). We have the following proposition on $\mathbb{E}(\gamma^\star_1)$.
\begin{proposition}\label{prop:bound_Rayleigh_fully}
For the case with Rayleigh fading channel, we have
\begin{equation}
\begin{split}
\left(\frac{\pi}{4}N^2+(M_r-1)N\right)\left(\frac{\pi}{4}N^2+(M_t-1)N\right)\\\le\mathbb{E}\left(\gamma^\star_1\right)\le \frac{\pi^2 M_tM_rN^4}{16}.
\end{split}
\end{equation}
\end{proposition}
\begin{IEEEproof}
See Appendix~\ref{sec:proof_of_proposition_ref_bound_rayleigh}.
\end{IEEEproof}

\begin{remark}\label{Re:Rayleigh_fully}
Based on Proposition \ref{prop:bound_Rayleigh_fully}, it follows that for the case with the Rayleigh fading channel and with optimal joint beamforming (i.e., the optimal reflective beamforming in (P2) together with the MRT in Section III-A), the resultant average SNR with fully-passive IRS increases proportional to $N^4$.
\end{remark}

Next, we consider the reflective beamforming with semi-passive IRS. In this case, maximizing $\overline{\text{SNR}}_2(\mathbf{\Phi})$ is equivalent to maximizing $\|\hat{\mathbf G}_t^T\mathbf{\Phi}^T\mathbf a\|^2$, which is formulated as
\begin{subequations}
  \begin{align}\notag
    \text{(P3)}:\max_{\mathbf \Phi}&\quad  \|\hat{\mathbf G}_t^T\mathbf{\Phi}^T\mathbf a\|^2\\ \label{eq:unit-modulus_2} 
    \text { s.t. }& \quad  |\mathbf \Phi_{n,n}|=1, \forall n\in \mathcal{N}.
  \end{align}
\end{subequations}
Problem (P3) is non-convex due to the unit-modulus constraint in \eqref{eq:unit-modulus_2}, which has a similar structure as the SNR maximization problem in IRS-assisted wireless communication system \cite{8811733} and can be solved using SDR together with Gaussian randomization, for which the details are omitted for brevity.



Based on the optimization in (P3), we analyze the resultant sensing SNR performance with semi-passive IRS. Let $\gamma^\star_2$ denote the optimal value of problem (P3). We have  the following proposition on  $\mathbb{E}(\gamma^\star_2)$.

\begin{proposition}\label{prop:bound_Rayleigh}
For the case with Rayleigh fading channel, we have
\begin{equation}\label{eq:SNR_bound_semi}
M_t\left(\frac{\pi}{4}N^2+(M_r-1)N\right)
 \le\mathbb{E}\left(\gamma^\star_2\right)
  \le \frac{\pi M_t M_rN^2}{4}.
\end{equation}
\end{proposition}
\begin{IEEEproof}
The proof is simliar as that for Proposition~\ref{prop:bound_Rayleigh_fully}, which is omitted for brevity.
\end{IEEEproof}
\begin{remark}\label{Re:Rayleigh_semi}
Based on Proposition \ref{prop:bound_Rayleigh}, it follows that for the case with the Rayleigh fading channel and with the optimal joint beamforming (i.e., the optimal reflective beamforming in (P3) together with the maximum ratio transmission in Section III-A), the resultant average SNR with semi-passive IRS increases proportional to $N^2$.
\end{remark}





By combining Propositions~\ref{prop:bound_Rayleigh_fully} and \ref{prop:bound_Rayleigh}, we have the following proposition.
\begin{proposition} \label{lemma_Rayleigh}
For the case with Rayleigh fading channel and with optimal joint beamforming, the average sensing SNR with fully-passive IRS is greater than that with semi-passive IRS (i.e., $\mathbb{E}(\overline{\text{SNR}}_1(\mathbf{\Phi}))>\mathbb{E}(\overline{\text{SNR}}_2(\mathbf{\Phi}))$) when 
\begin{equation}
N>2\sqrt{\frac{M_tM_r}{\pi L(d)}}-\frac{4}{\pi}\left(\max(M_t,M_r)-1\right).
\end{equation}

\end{proposition}

\section{Numerical Results}\label{sec:numerical_results}

This section provides numerical results to evaluate the sensing SNR performance of the fully-passive and semi-passive IRS sensing systems. The distance-dependent path loss is modeled as $L(d)=K_0\left(\frac{d}{d_0}\right)^{-\alpha_0}$,
where $d$ is the distance of the transmission link, $K_0=-30~ \text{dB}$ is the path loss at the reference distance $d_0=1~ \text{m}$, and the path-loss exponent $\alpha_0$ is set as $2.2$ and $2.0$ for the BS-IRS and IRS-target links, respectively. The BS, the IRS, and the target are located at coordinate $(0,0)$, $(1~\text{m},1~\text{m})$, and $(1~\text{m},-5~\text{m})$, respectively.  We also set $M_t=M_r=5$, $T=256$, $P_0 =30~\text{dBm}$, and $\sigma^2 = -90~\text{dBm}$. In the simulation, for the Rayleigh fading channel case, the simulation results are obtained by averaging over $100$ independent realizations. 

For performance comparison, we also consider the following benchmark schemes for transmit and reflective beamforming designs, in addition to the joint beamforming (BF) design in Section III.

\subsubsection{Reflective beamforming only with isotropic transmission (Reflective BF only)} The BS uses the isotropic transmission by setting the sample covariance matrix as $\mathbf R = P_0/M_t\mathbf I_{M_t}$. Then, the reflective beamforming at the IRS is optimized to maximize the sensing SNR.

\subsubsection{Transmit beamforming only with random reflection (Transmit BF only)} We consider the random reflecting phase shifts at the IRS, based on which the transmit beamforming at the BS is optimized based on MRT to maximize the sensing SNR.

\subsubsection{Isotropic transmission with random reflection (Without optimization)} We consider the isotropic transmission covariance matrix $\mathbf R = P_0/M_t\mathbf I_{M_t}$ at the BS and the  random reflecting phase shifts at the IRS.

\begin{figure}[t]
	\centering
	\centering
    \includegraphics[width=0.35\textwidth]{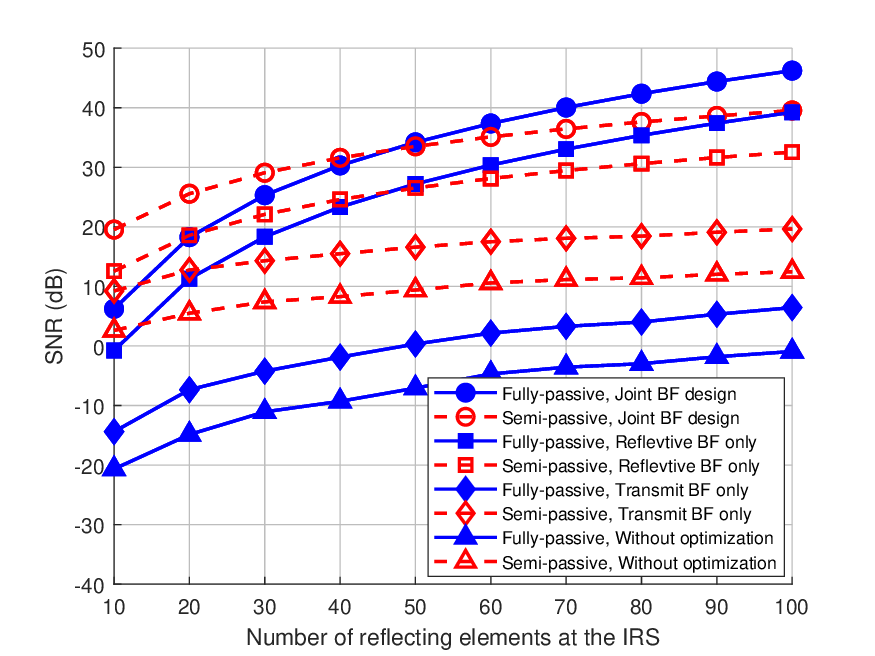}
    \caption{The SNR versus the number of reflecting elements at the IRS $N$ when the BS-IRS channels are LoS.}
    \label{SNR_N_LoS}
\end{figure}
\begin{figure}[t]
	\centering
    \includegraphics[width=0.35\textwidth]{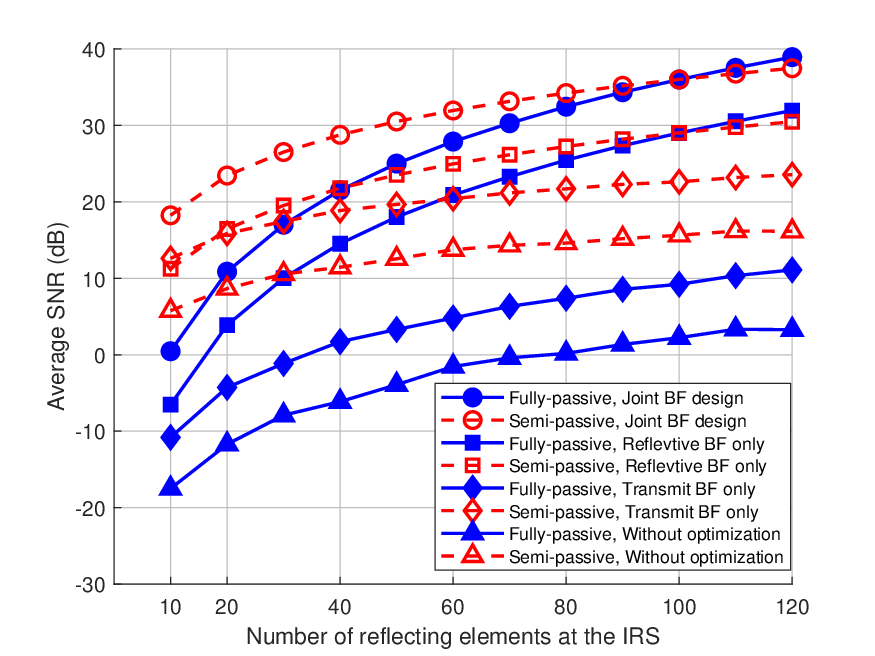}
    \caption{The average SNR versus the number of reflecting elements at the IRS $N$ when the BS-IRS channels follow Rayleigh fading.}
    \label{SNR_N_NLoS}
\end{figure}

Figs.~\ref{SNR_N_LoS} and \ref{SNR_N_NLoS} show the SNR versus the number of reflecting elements at IRS $N$ when  the BS-IRS channels are LoS and Rayleigh fading, respectively. For the two schemes with reflective beamforming optimization (i.e., the schemes with the joint BF design and reflective BF only), it is observed that when $N>46$ (or $N>100$) for the LoS channel (or Rayleigh fading channel), the sensing SNR of the fully-passive IRS sensing outperforms that of the semi-passive IRS sensing. This is due to the fact that in this case the additional reflective beamforming gain over the backward link from the IRS to the BS exceeds the corresponding path loss( see Propositions~\ref{lemma_LoS} and \ref{lemma_Rayleigh}). Meanwhile, by increasing $N$ from $10$ to $100$, the SNRs of the fully-passive and semi-passive IRS sensing systems increase $40$~dB and $20$~dB (or $38.47$~dB and $19.24$~dB) for the LoS channel (or Rayleigh fading channel), respectively. This is consistent with Remarks \ref{Re:Rayleigh_fully} and \ref{Re:Rayleigh_semi}. Finally, for the two benchmark schemes without reflective beamforming design (i.e., the schemes with the transmit BF only and that without optimization), it is observed that the SNR of the fully-passive IRS sensing is always less than that of the semi-passive IRS sensing in the whole region of $N$, due to the lack of reflective beamforming gains.

\section{Conclusion}
This paper analyzed the sensing SNR performance of the fully-passive and semi-passive IRS-enabled NLoS sensing systems with joint transmit and reflective beamforming optimization. It was shown that when the number of reflecting elements $N$ is sufficiently large, the sensing SNR increases proportional to $N^4$ and $N^2$ for the fully-passive and semi-passive IRS sensing systems, respectively. Thanks to the double reflective beamforming gains provided by the IRS over both the forward and backward links, the fully-passive IRS sensing outperforms the semi-passive counterpart when $N$ is sufficiently large. 

\appendix
\subsection{Algorithm for Solving Problem (P2)}\label{sub:algorithm_for_p2}
We solve problem (P2) by using the techniques of SDR and SCA. In particular, by letting $\mathbf v = [e^{j\phi_1},\cdots,e^{j\phi_{N}}]^{T}$ denote the vector collecting the $N$ reflecting coefficients at the IRS, we have 
\begin{equation}
\begin{split}
\|\hat{\mathbf G}_t^T\mathbf{\Phi}^T\mathbf a\|^2\|\hat{\mathbf G}_r\mathbf{\Phi}^T\mathbf a\|^2=&\|\hat{\mathbf G}_r \mathrm{diag}(\mathbf a) \mathbf v\|^2\|\hat{\mathbf G}_t^T\mathrm{diag}(\mathbf a) \mathbf v\|^2\\
=&\mathbf v^H\mathbf R_1\mathbf v \mathbf v^H\mathbf R_2\mathbf v,
\end{split}
\end{equation}
where $\mathbf R_1 = \mathrm{diag}(\mathbf a^*)\hat{\mathbf G}_r^H \hat{\mathbf G}_r \mathrm{diag}(\mathbf a)$ and $\mathbf R_2 = \mathrm{diag}(\mathbf a^*)\hat{\mathbf G}_t^* \hat{\mathbf G}_t^T \mathrm{diag}(\mathbf a)$. Next, we define $\mathbf V=\mathbf v\mathbf v^{H}$ with $\mathbf V \succeq \mathbf{0}$ and $\mathrm {rank}(\mathbf V)=1$. Problem (P2) is reformulated as 
\begin{subequations}
  \begin{align}\notag
    \text{(P2.1)}:\max_{\mathbf V\succeq \mathbf{0}}&\quad  \mathrm {tr}(\mathbf R_1 \mathbf V)\mathrm {tr}(\mathbf R_2 \mathbf V)\\ \label{eq:diagonal_one}
    \text { s.t. }& \quad  {\mathbf V_{n,n}}=1, \forall n\in \mathcal{N}\\\label{eq:rank-one}
    &\quad \mathrm {rank}(\mathbf V)=1. 
  \end{align}
\end{subequations}

By relaxing the rank-one constraint in \eqref{eq:rank-one}, the SDR version of problem (P2.1) is obtained as (SDR2.1).
Then, we use SCA to approximate the non-concave objective function in an iterative manner. In each iteration $r$, with local point $\mathbf V^{(r)}$, a global linear lower bound function of $\mathrm {tr}(\mathbf R_1 \mathbf V)\mathrm {tr}(\mathbf R_2 \mathbf V)$ is obtained by using its first-order Taylor expansion, i.e.,
\begin{equation}
\begin{split}
\mathrm {tr}(\mathbf R_1 \mathbf V)\mathrm {tr}(\mathbf R_2 \mathbf V)
\ge& \mathrm {tr}(\mathbf R_1 \mathbf V^{(r)})\mathrm {tr}(\mathbf R_2 \mathbf V^{(r)})\\
&+\mathrm {tr}(\mathbf R_1 \mathbf V^{(r)})\mathrm {tr}(\mathbf R_2 (\mathbf V-\mathbf V^{(r)}))\\
&+\mathrm {tr}(\mathbf R_1(\mathbf V-\mathbf V^{(r)}))\mathrm {tr}(\mathbf R_2 \mathbf V^{(r)})\\
\triangleq& f^{(r)}(\mathbf V).
\end{split}
\end{equation}

By substituting $\mathrm {tr}(\mathbf R_1 \mathbf V)\mathrm {tr}(\mathbf R_2 \mathbf V)$ with $f^{(r)}(\mathbf V)$ in iteration $r$, problem (SDR2.1) is approximated as a convex problem (SDR$2.1.r$), which can be optimally solved by CVX\cite{cvx}.
Let $\mathbf V^{(r,\star)}$ denote the optimal solution to problem (SDR$2.1.r$), which is then updated to be the local point $\mathbf V^{(r+1)}$ for the next inner iteration $r + 1$. As $f^{(r)}(\mathbf V)$ is a lower bound of $\mathrm {tr}(\mathbf R_1 \mathbf V)\mathrm {tr}(\mathbf R_2 \mathbf V)$, we have
$\mathrm {tr}(\mathbf R_1 \mathbf V^{(r+1)})\mathrm {tr}(\mathbf R_2 \mathbf V^{(r+1)})\ge f^{(r)}(\mathbf V^{(r+1)})
\ge f^{(r)}(\mathbf V^{(r)})
=\mathrm {tr}(\mathbf R_1 \mathbf V^{(r)})\mathrm {tr}(\mathbf R_2 \mathbf V^{(r)})$.
Thus, each iteration leads to a non-decreasing objective value for problem (SDR2.1). As a result, the convergence of SCA for solving problem (SDR2.1) is ensured. Let $\hat{\mathbf V}$ denote the obtained  solution to problem (SDR2.1) where $\mathrm{rank}(\hat{\mathbf V})>1$ may hold. 

Finally, we use Gaussian randomization to construct an approximate rank-one solution of $\mathbf V$ to problem (P2.1). First, we generate a number of randomizations $\mathbf r \sim \mathcal{CN}(\mathbf{0},\hat{\mathbf V})$, and then construct candidate solutions as $\mathbf{v}=e^{j\mathrm {arg}(\mathbf r)}$. The solution of (P2.1) is chosen from the candidate solutions as the one achieving the maximum objective value of (P2.1).

\addtolength{\topmargin}{0.01in}
\subsection{Proof of Proposition \ref{prop:bound_Rayleigh_fully}}\label{sec:proof_of_proposition_ref_bound_rayleigh}
First, we introduce the upper bound of $\mathbb{E}(\|\hat{\mathbf G}_r\mathbf{\Phi}^T\mathbf a\|^2\|\hat{\mathbf G}_t^T\mathbf{\Phi}^T\mathbf a\|^2)$. We define $\hat{\mathbf G}_t= [\mathbf g_{1}, \cdots,\mathbf g_{\min(M_t,M_r)},\cdots,\mathbf g_{M_t}]$, $\hat{\mathbf G}_r^T= [\mathbf g_{1}, \cdots,\mathbf g_{\min(M_t,M_r)},\cdots,\mathbf g_{M_r}]$, and $\mathbf c = \mathbf{\Phi}^{T}\mathbf a$. An upper bound of $\mathbb{E}(\|\hat{\mathbf G}_r\mathbf{\Phi}^T\mathbf a\|^2\|\hat{\mathbf G}_t^T\mathbf{\Phi}^T\mathbf a\|^2)$ is given as
\begin{equation}\notag
\begin{split}
&\mathbb{E}\left(\|\hat{\mathbf{G}}_t^{T}\mathbf{\Phi}^{T}\mathbf a\|^2\|\hat{\mathbf{G}}_r\mathbf{\Phi}^{T}\mathbf a\|^2\right)\!=\!  \mathbb{E}\left(\sum_{m=1}^{M_t}|\mathbf g_m^T\mathbf c|^2\sum_{m=1}^{M_r}|\mathbf g_m^T\mathbf c|^2\right)\\
&\stackrel{(a_{1})}{\le} \mathbb{E}\left(\sum_{m=1}^{M_t}\left(\sum_{n=1}^N|g_{n,m}|\right)^2\sum_{m=1}^{M_r}\left(\sum_{n=1}^N|g_{n,m}|\right)^2\right)\\
&\stackrel{(a_{2})}{=}\frac{\pi}{16}M_tM_rN^4,
\end{split}
\end{equation}
where inequality $(a_{1})$ holds due to $|c_n|=1$ with $c_n$ denoting the $n$-th element of $\mathbf c$, and equality $(a_{2})$ holds due to $\|\mathbf c\|^2=N$ and $\mathbb{E}(|g_{n,m}|)=\sqrt{\pi}/2$ with $g_{n,m}$ denoting the $n$-the element of $\mathbf g_{m}$. 

Next, we consider a special case of reflective beamforming design to give a lower bound of $\mathbb{E}(\|\hat{\mathbf G}_r\mathbf{\Phi}^T\mathbf a\|^2\|\hat{\mathbf G}_t^T\mathbf{\Phi}^T\mathbf a\|^2)$. Towards this end, we choose any $i$ with $1\le i\le\min(M_t,M_r)$ and accordingly set $\hat{\mathbf{\Phi}} = \mathrm {diag}(e^{j\hat{\phi}_1},\ldots,e^{j\hat{\phi}_{N}})$ with
$\hat{\phi}_n = -\mathrm{arg}(g_{n,i}) -\mathrm{arg}(a_n), \forall n \in \mathcal{N}$, and $\hat{\mathbf c}=\hat{\mathbf{\Phi}}^T\mathbf a$.
In this case, we have
\begin{equation}\notag
\begin{split}
&\mathbb{E}\left(\|\hat{\mathbf{G}}_t^{T}\hat{\mathbf{\Phi}}^{T}\mathbf a\|^2\|\hat{\mathbf{G}}_r\hat{\mathbf{\Phi}}^{T}\mathbf a\|^2\right)\\
\!=&\!  \mathbb{E}\left(\left(|\mathbf g_i^T\hat{\mathbf c}|^2\!+\!\sum_{m=1,m\neq i}^{M_t}\!|\mathbf g_m^T\hat{\mathbf c}|^2\right)\right.\\
&\quad\left.\left(|\mathbf g_i^T\hat{\mathbf c}|^2+\sum_{m=1,m\neq i}^{M_r}|\mathbf g_m^T\hat{\mathbf c}|^2\right)\right)\\
&=\left(\frac{\pi}{4}N^2+(M_t-1)N\right)\left(\frac{\pi}{4}N^2+(M_r-1)N\right),
\end{split}
\end{equation}
which serves as a lower bound of $\mathbb{E}(\|\hat{\mathbf G}_r\mathbf{\Phi}^T\mathbf a\|^2\|\hat{\mathbf G}_t^T\mathbf{\Phi}^T\mathbf a\|^2)$.


\ifCLASSOPTIONcaptionsoff
  \newpage
\fi

\bibliographystyle{IEEEtran}
\bibliography{IEEEabrv,mybibfile}

\end{document}